# Comparative Study of Open-loop Transmit Diversity Schemes for Four Transmit Antennas in Coded OFDM Systems

Chau Yuen, Yan Wu, and Sumei Sun
Institute for Infocomm Research (I²R), Singapore
{cyuen, wuyan, sunsm}@i2r.a-star.edu.sg

*Abstract:* **We compare four open-loop transmit diversity schemes in a coded Orthogonal Frequency Division Multiplexing (OFDM) system with four transmit antennas, namely cyclic delay diversity (CDD), Space-Time Block Code (STBC, Alamouti code is used) with CDD, Quasi-Orthogonal STBC (QO-STBC) and Minimum-Decoding-Complexity QOSTBC (MDC-QOSTBC). We show that in a coded system with low code rate, a scheme with spatial transmit diversity of second order can achieve similar performance to that with spatial transmit diversity of fourth order due to the additional diversity provided by the phase shift diversity with channel coding. In addition, we also compare the decoding complexity and other features of the above four mentioned schemes, such as the requirement for the training signals, hybrid automatic retransmission request (HARQ), etc. The discussions in this paper can be readily applied to future wireless communication systems, such as mobile systems beyond 3G, IEEE 802.11 wireless LAN, or IEEE 802.16 WiMAX, that employ more than two transmit antennas and OFDM.**

## I. INTRODUCTION

We consider a multiple-input multiple-output (MIMO) system with four transmit antennas at the base station. Since the wireless channels experience fading, transmit diversity plays an important role. In this paper, we compare four simple transmit diversity schemes in a coded Orthogonal Frequency Division Multiplexing (OFDM) system.

The first scheme is cyclic delay diversity (CDD) [1], also known as cyclic shift diversity CSD. Since it can be treated as phase diversity in frequency domain, it does not provide any spatial diversity, and relies much on the capability of the channel coding. The second scheme is the combination of Space-Time Block Code (STBC) with CDD [2]. We use the rate-1 orthogonal STBC, namely Alamouti STBC, which is originally designed for two transmit antennas, and combine it with CDD to support four transmit antennas. In this case, it can provide a spatial diversity of two and yet achieve maximum-likelihood detection (MLD) with linear complexity.

As no orthogonal design can achieve full rate when there are four transmit antennas, we consider two rate-1 non-orthogonal STBCs that can provide spatial transmit diversity of level four, they are Quasi-Orthogonal STBC (QO-STBC) [3] and Minimum-Decoding-Complexity QO-STBC (MDC-QOSTBC) [4]. These STBCs are selected as they are "quasi-orthogonal" and hence have a lower decoding complexity than other STBC schemes for four transmit antennas. The MLD decoding search space for the above mentioned schemes is given in Table 1.

As shown in Table 1, for a constellation of size-$M$, an orthogonal design only requires a search space of sqrt($M$), while QO-STBC requires a search space of $M^2$ and MDC-QOSTBC requires a search space of $M$. Although MDC-QOSTBC has a slightly higher complexity than the orthogonal design, but such complexity is still manageable in practical system. And this is the advantage of MDC-QOSTBC over QO-STBC.

Table 1 MLD search space for QO-STBC and MDC-QOSTBC

|  | Decoding search space | | |
| --- | --- | --- | --- |
|  | QO-STBC | MDC-QOSTBC | Alamouti, CDD or Alamouti+CDD |
| QPSK | 16 | 4 | 2 |
| 16QAM | 256 | 16 | 4 |
| $M$ points | $M^2$ | $M$ | sqrt($M$) |

In the rest of the paper, we will first discuss the soft decision decoding of MDC-QOSTBC in a coded system. We then compare the decoding performance of the four transmit diversity schemes in a coded OFDM system, and discuss on the features and merits of the schemes respectively.

## II. MDC-QOSTBC IN CODED SYSTEM

The coded performance of QO-STBC in an OFDM system has been reported in [5][6]. However, the coded performance of MDC-QOSTBC has yet been reported in the literature. Consider the MDC-QOSTBC as shown below:

$$C = \begin{bmatrix} x_1 & x_2 & x_3 & x_4 \\ -x_2^* & x_1^* & -x_4^* & x_3^* \\ x_3 & x_4 & x_1 & x_2 \\ -x_4^* & x_3^* & -x_2^* & x_1^* \end{bmatrix} \quad (1)$$

where $x_1 = c_1^R + jc_3^R$, $x_2 = c_2^R + jc_4^R$, $x_3 = -c_1^I + jc_3^I$, $x_4 = -c_2^I + jc_4^I$, and $c_i = c_i^R + jc_i^I$ ($1 \le i \le 4$) are the transmitted data symbols, while $c_i^R$ and $c_i^I$ are the real and imaginary parts of a complex symbol.

The received signals can be written as:

$$\begin{aligned} \mathbf{r} &= \mathbf{Ch} + \mathbf{n} \\ &= \mathbf{H}_{eq}\mathbf{c} + \mathbf{n} \end{aligned} \quad (2)$$

where $\mathbf{H}_{eq}$ is the equivalent channel as described in [10], and $\mathbf{c}$ is the real-valued transmitted signal, i.e., $\mathbf{c} = \begin{bmatrix} c_1^R & c_1^I & c_2^R & c_2^I & c_3^R & c_3^I & c_4^R & c_4^I \end{bmatrix}^T$.

By applying the linear matched filter $\mathbf{H}_{eq}^*$ and whitening filter $\mathbf{H}_w$ to (2) as described in [9], we get:

$$\begin{aligned} \mathbf{H}_w \mathbf{H}_{eq}^* \mathbf{r} &= \mathbf{H}_w \mathbf{H}_{eq}^* \mathbf{Hc} + \mathbf{H}_w \mathbf{H}_{eq}^* \mathbf{n} \\ &= \mathbf{H}_{final}\mathbf{c} + \tilde{\mathbf{n}} \end{aligned} \quad (3)$$

where $\tilde{\mathbf{n}}$ is white noise.

It can be easily shown that $\mathbf{H}_{final}$ is a block diagonal matrix, with four 2-by-2 submatrices. That is, the four transmitted symbols are separated into four orthogonal groups, each of them can be decoded independently. We can represent the first group as follows:

$$\begin{bmatrix} y_1 \\ y_2 \end{bmatrix} = \begin{bmatrix} h_{11} & h_{12} \\ h_{21} & h_{22} \end{bmatrix} \begin{bmatrix} c_1^R \\ c_1^I \end{bmatrix} + \begin{bmatrix} v_1 \\ v_2 \end{bmatrix} \Rightarrow \mathbf{y} = \mathbf{Hc} + \mathbf{v} \quad (4)$$

where $v_1$ and $v_2$ are AGWN noise, and $y_1$ and $y_2$ are the output of the matched and whitening filter. So the MLD can be performed symbol-by-symbol independently.

Let's assume that each of the symbols is QPSK, hence the real and imaginary part can only be the value of 1 or -1. The log-likelihood ratio for data bit $b_1$ can be computed as:

$$\begin{aligned} b_1 &= \log \frac{p(c_1^R = 1 | \mathbf{y})}{p(c_1^R = -1 | \mathbf{y})} = \log \frac{p(\mathbf{y} | c_1^R = 1) p(c_1^R = 1)}{p(\mathbf{y} | c_1^R = -1) p(c_1^R = -1)} \\ &= \log \frac{p(\mathbf{y} | c_1^R = 1, c_1^I = 1) + p(\mathbf{y} | c_1^R = 1, c_1^I = -1)}{p(\mathbf{y} | c_1^R = -1, c_1^I = 1) + p(\mathbf{y} | c_1^R = -1, c_1^I = -1)} \end{aligned} \quad (5)$$

if we assume equal *a priori* probability for bits $c_1^R = 1$, and $c_1^R = -1$. Likewise the soft decision metric for the second bit can be computed accordingly.

### III. SIMULATION RESULTS

In this section, we present our performance evaluation results of the four transmit diversity schemes. We consider a MIMO system with four transmit and two receive antennas. For error control coding, we employ the turbo codes from the UMTS standard with feedforward polynomial $1+D+D^3$, and feedback polynomial $1+D^2+D^3$. Information code block length is 594 bits for rate-1/2 and 1056 bits for rate-8/9. For decoding, Max-Log-Map with 8 iterations is implemented. We use the TU6 channel and assume that the channel is spatially-uncorrelated and perfectly known at the receiver. The cyclic delay values are [0 64 128 192] respectively for each of the transmit antennas for CDD schemes. There are 512 subcarriers per OFDM symbol. We will compare the following four transmit diversity schemes, all for four transmit antennas:

- CDD
- Alamouti + CDD
- QO-STBC
- MDC-QOSTBC

For the decoding, LMMSE receiver is used with QOSTBC while MLD for the rest of the scheme.

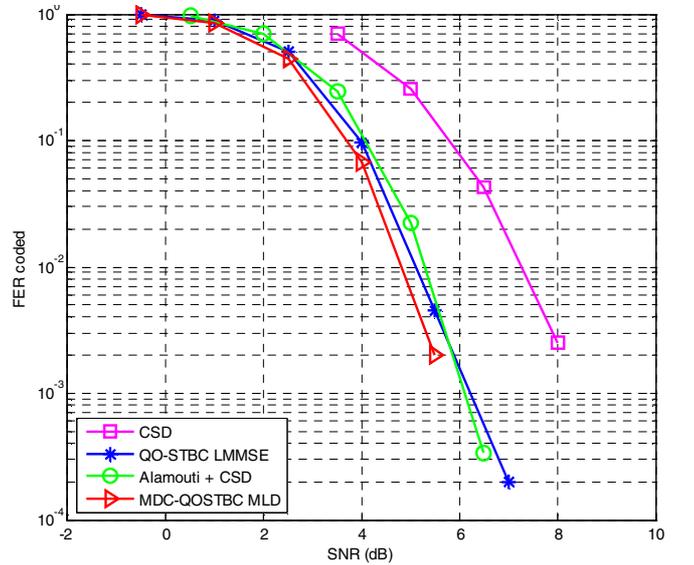

Figure 1 Simulated FER for 4tx-2rx, QPSK with turbo code rate-8/9

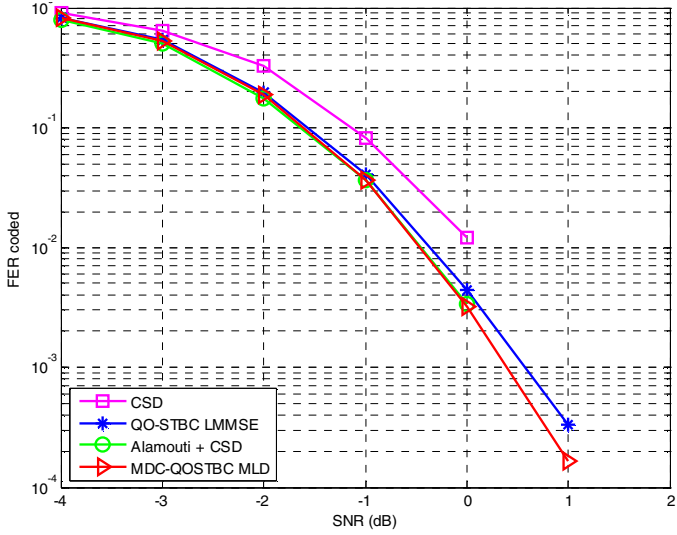

Figure 2 Simulated FER for 4tx-2rx, QPSK with turbo code rate-1/2

The simulations results with four transmit and two receive antennas system with QPSK modulation are shown in Figure 1 for rate-8/9 and Figure 2 for rate/1/2 turbo code, respectively. Observations can be summarized as follows:

- MDC-QOSTBC and Alamouti+CDD outperform CDD by at least 0.5dB (at coded FER $10^{-1}$ or below) in all cases. The gap with CDD is larger when the code rate is high, as CDD mainly obtain the diversity from the channel coding, hence when the code rate is high (e.g. for the data channel), CDD will perform poorly.

- MDC-QOSTBC performs the best when the code rate is high. This is mainly because MDC-QOSTBC obtains most of the transmit diversity from its code structure instead of from the channel coding.

- MDC-QOSTBC with MLD has a better performance than QO-STBC with LMMSE. The low search space feature of MDC-QOSTBC makes MLD possible, and this is the advantage over QO-STBC.

- Though it is not shown in the figure, MDC-QOSTBC has the same performance as QO-STBC when LMMSE is used [9].

To summarize, in terms of performance, Alamouti+CDD and MDC-QOSTBC are the two best schemes. And MDC-QOSTBC performs the best in all sorts of conditions that we have studied. In the next section, we will discuss additional features of MDC-QOSTBC to make it more interesting for practical usage.

## IV. ADDITIONAL FEATURES

By rewriting the codeword of MDC-QOSTBC in (1) into (6), the MDC-QOSTBC consists of many other schemes as a special case, such as:

- a rate-2 transmit diversity-2 code for four transmit antennas **DSTTD** [8]
- a rate-4 spatial multiplexing **SM** for four transmit antennas [1]
- a rate-2 full transmit diversity code for two transmit antennas **XTD** [7]

$$\mathbf{C} = \begin{bmatrix} c_1^R + jc_3^R & c_2^R + jc_4^R & -c_1^I + jc_3^I & -c_2^I + jc_4^I \\ -c_2^R + jc_4^R & c_1^R - jc_3^R & c_2^I - jc_4^I & -c_1^I - jc_3^I \\ -c_1^I + jc_3^I & -c_2^I + jc_4^I & c_1^R + jc_3^R & c_2^R + jc_4^R \\ c_2^I + jc_4^I & -c_1^I - jc_3^I & -c_2^R + jc_4^R & c_1^R - jc_3^R \end{bmatrix}$$

$$= \begin{bmatrix} x_1 & x_2 & x_3 & x_4 \\ -x_2^* & x_1^* & -x_4^* & x_3^* \\ x_3 & x_4 & x_1 & x_2 \\ -x_4^* & x_3^* & -x_2^* & x_1^* \end{bmatrix} \quad (6)$$

(XTD, SM, D-STTD annotations on rows)

This feature can be useful in HARQ [8] and may lead to simplified receiver design. First of all, in order to achieve maximum throughput, the system can use **SM** scheme by transmitting the first row of **C**. If such transmission leads to detection error, the 2nd row of **C** can be transmitted, and the receiver can then combine this received signal with the one received previously and decode them as **DSTTD** code. By doing so, a transmit diversity of level two can be achieved. If such transmission still has error, the 3rd and 4th row of **C** can be transmitted, and this is equivalent to transmitting the rate-1 **MDC-QOSTBC**, and the receiver can then combine all the received signals, and perform a ML decoding. This results in a transmission scheme with transmit diversity of level four. Hence such HARQ scheme has the ability to increase the transmit diversity by lowering the transmission rate, and at the same time, makes full use of the previous transmission rather than discard them.

It can be noticed that **XTD** is a special case for MDC-QOSTBC, which suggest possible simplification in the receiver design. In addition, it also posts an interesting area in antenna selection, which we leave it for future study.

By properly designing the reference signaling, CDD and Alamouti+CDD can appear to be transparent to the receiver, i.e. the receiver sees a single stream or Alamouti transmission without knowing existence of CDD. Unfortunately, such feature is not available for MDC-QOSTBC.

A summary on the comparisons of CDD, Alamouti+CDD and MDC-QOSTBC is shown in Table 2.

Table 2 Comparison btw different transmit diversity schemes

|  | CDD | Alamouti + CDD | MDC-QOSTBC |
|---|---|---|---|
| Decoding performance | X | √ | √ |
| Transparent to the receiver (depending on the ref. signal) | √ | √ | X |
| Not sensitive to code rates and channel multipath condition | X | X | √ |
| Others:<br>• Part of the HARQ scheme as described in [8].<br>• Include other STBC, e.g. XTD as a special case. | X | X | √ |

## V. CONCLUSION

We first present the decoding of MDC-QOSTBC in a coded system. We show that in a coded OFDM system, a transmit diversity scheme with only spatial transmit diversity of level two can perform as well as a scheme with spatial transmit diversity of level four. This is due to the additional diversity provided by the channel coding. Hence when the channel coding is strong (for example for the case of control channel), Alamouti with CDD seems to be the best candidates; while when the channel coding is weak (for example for the case of data channel), MDC-QSTBC seems to be the best candidates. In addition, CDD scheme has the advantage of being transparent to the receiver by properly design the reference signal (i.e. pilot), MDC-QOSTBC has the advantage of be part of an interesting hybrid ARQ.

## REFERENCES

[1] S. Kaiser, ''Spatial transmit diversity techniques for broadband OFDM systems,'' in *IEEE Global TeleCommunications Conference*, vol. 3, pp. 1824--1828, Nov. 2000

[2] ETRI, "Combined STBC/CDD transmission scheme for multiple antennas", R1-060438, *3GPP LTE*, Feb 2006.

[3] H. Jafarkhani, "A quasi-orthogonal space-time block code", *IEEE Trans. on Communications,* vol. 49, pp. 1-4, Jan. 2001.

[4] C. Yuen, Y. L. Guan, and T. T. Tjhung, "Quasi-orthogonal STBC with minimum decoding complexity", *IEEE Trans. Wireless Comms.*, vol. 4, Sept. 2005, pp. 2089 – 2094.

[5] C. K. Sung, J. Kim, and I. Lee, "Quasi-orthogonal STBC with iterative decoding in bit interleaved coded modulation", *IEEE VTC 2004*, pp. 1323-1327.

[6] J. Kim and I. Lee, "Space-time coded OFDM systems with four transmit antennas", *IEEE VTC 2004*, pp. 2434-2438.

[7] LG Electronics, "High-rate STC for open-loop MIMO with 2tx antennas", R1-060967, *3GPP LTE*, March 2006.

[8] Nortel, "QO-STFBC / Double-STTD based H-ARQ for four transmit antennas", R1-060149, *3GPP LTE*, Jan 2006.

[9] C. Yuen; Y. L. Guan; T. T. Tjhung, "Decoding Of Quasi-Orthogonal Space-Time Block Code With Noise Whitening", *IEEE PIMRC*, Beijing, China, Sept 2003. Volume 3, pp. 2166 – 2170.

[10] B. Hassibi and B. M. Hochwald, "High-rate codes that are linear in space and time", *IEEE Trans. on Information Theory*, vol. 48, pp. 1804 –1824, Jul. 2002.